\documentclass[aps,pre,twocolumn,groupedaddress,showpacs]{revtex4}
\usepackage{amssymb}
\usepackage{graphicx,color}
\definecolor{r}{rgb}{1,0,0}
\definecolor{g}{rgb}{0,1,0}
\definecolor{b}{rgb}{0,0,1}

\begin{document}


\title{Kinetics of Gravity-Driven Water Channels Under Steady Rainfall}



\author{Cesare M. Cejas$^{1}$, Yuli Wei$^{1,2}$, Remi Barrois$^{1}$, 
Christian Fr$\acute{e}$tigny$^{3}$, Douglas J. Durian$^2$, and R$\acute{e}$mi Dreyfus$^{1}$}
\affiliation{$^{1}$Complex Assemblies of Soft Matter, CNRS-Solvay-UPenn UMI 3254, Bristol, PA 19007-3624, USA}
\affiliation{$^{2}$Department of Physics and Astronomy, University of Pennsylvania, Philadelphia, PA 19104-6396, USA}
\affiliation{$^{3}$Physico-chimie des Polym$\grave{e}$res et des Milieux Dispers$\acute{e}$s CNRS PPMD UMR 7615 ESPCI, Paris, France 75005}


\date{\today}

\begin{abstract}
We investigate the formation of fingered flow in dry granular media under simulated rainfall using a quasi-2D experimental set-up composed of a random close packing of mono-disperse glass beads. Using controlled experiments, we analyze the finger instabilities that develop from the wetting front as a function of fundamental granular (particle size) and fluid properties (rainfall, viscosity).These finger instabilities act as precursors for water channels, which serve as outlets for water drainage. We look into the characteristics of the homogeneous wetting front and channel size as well as estimate relevant time scales involved in the instability formation and the velocity of the channel finger tip. We compare our experimental results with that of the well-known prediction developed by Parlange and Hill [1976]. This model is based on linear stability analysis of the growth of perturbations arising at the interface between two immiscible fluids. Results show that in terms of morphology, experiments agree with the proposed model. However, in terms of kinetics we nevertheless account for another term that describes the homogenization of the wetting front. This result shows that the manner we introduce the fluid to a porous medium can also influence the formation of finger instabilities.
\end{abstract}

\pacs{to be determined}
%


\maketitle



Water infiltration in soil is a long-standing research topic due to a wealth of interesting physical phenomena, such as fluid-granular interactions, as well as also having a wide variety of industrial applications. For example, rainwater can induce solute leaching as it drives contaminants from the unsaturated zone just below the soil surface to deeper areas underground such as the water table~\cite{DeRooij00}. This affects the quality of groundwater and thus such infiltration studies have aimed to limit the adverse effects of groundwater contamination~\cite{Glass91, Wang98, Hillel03}. Both laboratory~\cite{Hill72, Diment85} and real field experiments~\cite{Ritsema98, Hendrickx93} have confirmed the existence of preferential drainage paths in sandy soils under uniform flow via rainfall or irrigation water. In agricultural applications, when water drains through preferential channels, drainage greatly reduces the quantity of water around the root zone that could otherwise be absorbed by the plants. Understanding the physical mechanisms involved in water infiltration during rain can help in developing novel techniques that could potentially have direct applications in soil remediation and water retention. Since infiltration is an example of multiphase flow, basic interest on the subject have initially focused on the dynamics of the interface between two immiscible fluids. 

Infiltration proceeds via the formation of preferential paths. Extensive experiments have shown that apart from soil structural heterogeneities like macropores~\cite{Davidson84}, preferential paths may also occur in homogeneous dry sand. This is due to the fingering instabilities developing from the interface of a wetting front that occur during initially uniform and gravity-driven fluid flow~\cite{Hill72, Diment85, Glass89, Baker90}. This has been observed in homogeneous sandy soil~\cite{Raats73, Vanommen89, Ritsema93, Yao93} but is nevertheless also proven in materials of varying wettability~\cite{Bond69, Wang00, Cejas13}.

Over the years, infiltration studies have employed empirical~\cite{Green11, Horton40, Chen06}, numerical~\cite{Brutsaert77, Pachepsky03}, and theoretical~\cite{Parlange76} solutions to describe the phenomenon observed in both real soil fields and laboratory simulations. Recent studies~\cite{Cueto-Felgueroso08, Cueto-Felgueroso09} have brought additional insight into existing equations in modelling gravity-driven flow. Such equations are fundamentally based on Richards$^{\prime}$ equation for unsaturated flow, which couples Darcy$^{\prime}$s law and mass conservation law. However, Richards$^{\prime}$ equation is unable to simulate fingering phenomenon~\cite{Cueto-Felgueroso08, Cueto-Felgueroso09, Nieber05}, thus extensions are normally added to account for certain aspects of multiphase flow~\cite{Eliassi01}. Previous studies~\cite{Glass91, Diment85, Raats73, Parlange76, Saffman58, Chuoke59} have proposed models to explain experiments based on parameters that condition wetting front instability, such as water repellency~\cite{Ritsema93, Ritsema98, Wang98} and water redistribution~\cite{Philip69}. However, to our knowledge, most studies on infiltration have focused on morphology of the water channels that form during infiltration. Also, not much has been performed with regards to understanding kinetics.  Some studies have focused on the change in the pressure jump that accompanies flow velocity through the unsaturated zone~\cite{DiCarlo00}. Others have focused on how flow velocity is affected by hydraulic properties such as conductivity and saturation~\cite{Raats73}. Still, not much has been brought to light regarding the influence of the water source, which is normally introduced to a porous medium in a homogeneous manner. Acquiring a full grasp of the dynamics of the phenomenon first requires comprehensive analysis of the fundamental physical features that arise from the infiltration process. This means that understanding how channel size and channel velocity are affected by granular and fluid properties remain to be key pieces in mapping out the entire puzzle of the phenomenon of finger instability.

In this paper, we present an experimental kinetic study on the dynamics of the formation of water channels during steady rainfall. Using a quasi-two dimensional (2D) set-up that simulates different rainfall rates and at the same time provides good visualization of water channel formation, we determine systematically the influence of physical parameters on the formation of the wetting front, instability, and propagation of water channels. These physical parameters include granular properties, such as particle size, which have been commonly studied. We also vary fluid viscosity, an important parameter whose effect on channel formation has not yet been sufficiently surveyed. Moreover, we also estimate relevant time scales involved in water channel formation, thereby providing additional information on the kinetics of the instability.


\section{Experiment}

We use a quasi-two-dimensional ($2$D) cell, of cross-sectional area $A$ (length, $l$ = $30$~cm, cell thickness, $e$ = $0.8$~cm). To reduce wetting effects on the glass wall, the sample cell is made hydrophobic by washing with hydrophobic silane solution (OMS Chemicals). We attach screen meshes at the bottom portion of the cell to freely circulate air and to freely drain water while preventing glass beads from emptying out of the 2D cell.

This cell is filled with random close packing of monodisperse glass beads (A-series, Potters Industries, Inc.) as our model soil system.  The glass beads are hydrophilic. We clean them by burning them in a furnace for $72$ hours at high temperature. Then we soak the glass beads in $1$~M HCl, rinse with deionized water, and then bake them in a vacuum oven for $12$ hours at $110^{\circ}$C and then left to cool at room temperature. Contact angle measurements on the clean hydrophilic beads reveal a contact angle, $\theta^*$, of $\theta^*$ $=$ $16\pm 2^{\circ}$. The packing porosity is $\epsilon$ = $0.36 - 0.40$ and is measured using the imbibition method. The cell is first filled with dry glass beads then the glass beads are taken out of the cell. Water is then slowly poured into the glass beads until they are fully saturated. The glass beads are weighed before and after the imbibition and since the density is known, the difference gives the pore volume. Calculating pore volume with respect to bulk volume of the cell gives the porosity.

We build a rain source with equally spaced glass capillary tubes (borosilicate micropipettes, VWR). The spacing between the tubes is $1$~cm. The rain source provides a constant rain rate $Q$ and we control the distance, $h$, between the tip of the capillary and the soil surface to control the droplet impinging speed. From the average masses of the raindrops, we estimate the droplet diameter to be $3$~mm. We suspend the sample cell beneath the rain source as shown in Fig. 1. 

\begin{figure}
\includegraphics[width=3.7in]{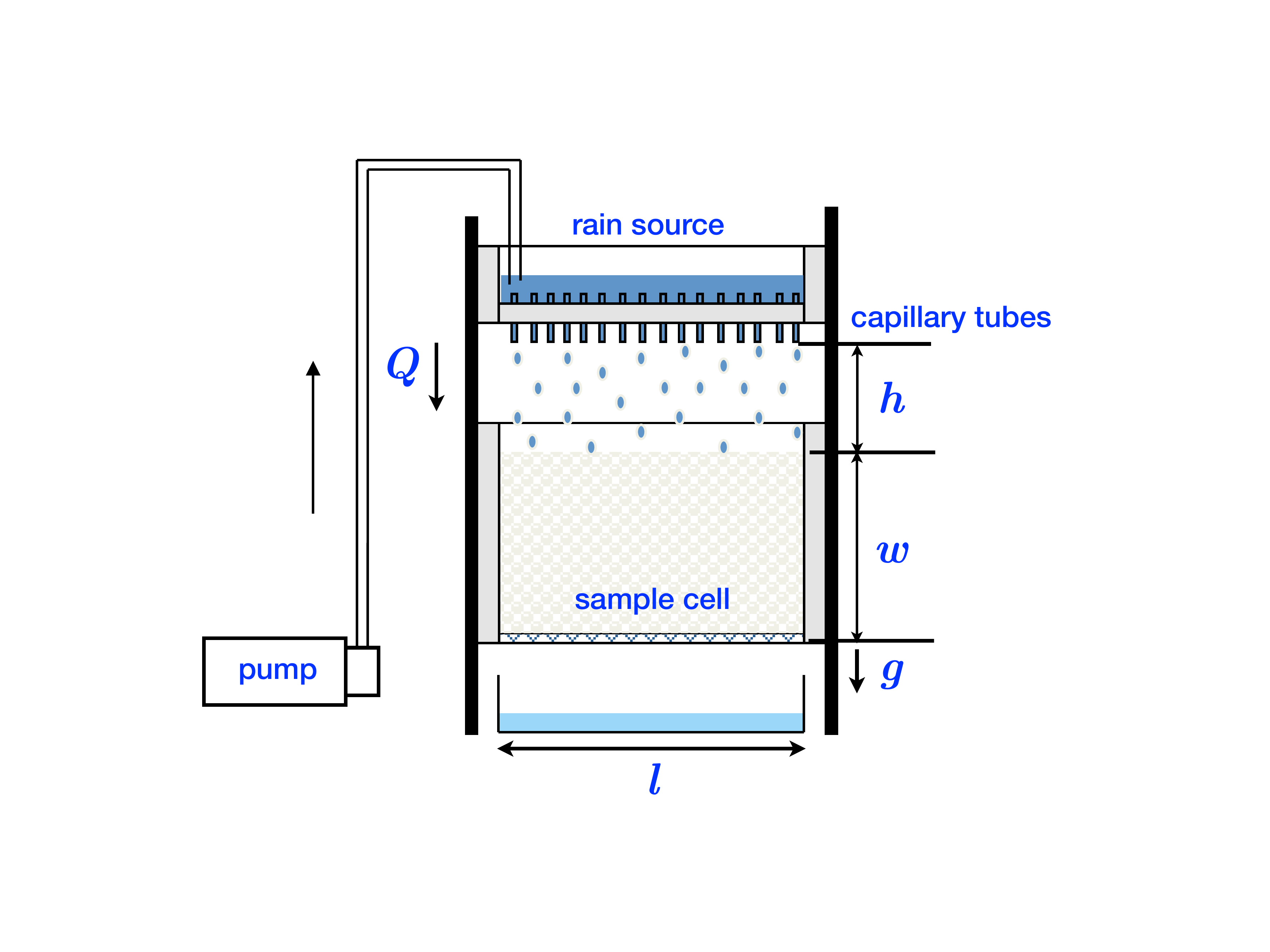}
\caption{(Color online) Diagram of the quasi-$2$D experimental set-up used to visualize the formation of water channels during steady rain. A rain source built with equally spaced capillary tubes provides a constant rainfall rate on the sample cell of cross-section $A$. The sample cell is filled with monodisperse glass beads as model soil and is suspended under the rain source. We can vary the distance between the capillary and model soil surface to vary the free-fall height of the raindrops.  }
 \label{Samplecell}
\end{figure}

We measure the rain rate by determining the volume of water per time per cross-sectional area. As expected, experiments show that the rain rate is proportional to the water level in the rain source. Because also of the design limitation of the size of the rain source, extremely high flow rates can only be achieved when the size of the capillary tubes is also modified. Control of the flow rate is set according to the water level height in the rain source and the size of the capillary tubes. Thus, to achieve higher flow rates, we vary the capillary tubes using readily available capillary tubes ($\pm0.5\%$) in the market: $5$~${\mu}$L, $10$~${\mu}$L, $25$~${\mu}$L, and $50$~${\mu}$L. The full lengths of all these commercial capillary tubes are $12.70 \pm 0.05$~cm and the outer diameters (OD) of all the tubes are measured to be within the range of approximately $1.6 -1.8$~$\pm$~$0.5$~mm. Since these tubes are in fact micropipettes, the volumes are calibrated only up to a certain effective length, which is $7.30 \pm 0.05$~cm. The inner diameters (ID) vary according to its volumetric capacity and can be calculated from the effective length. The values are $0.295 \pm 0.001$~mm, $0.418 \pm 0.002$~mm, $0.660 \pm 0.003$~mm, $0.934 \pm 0.004$~mm for the four aforementioned tubes respectively. But even though the capacity of the capillary and their ID values change, the OD values are roughly constant. Hence, the size of the droplet also roughly remains the same. 

The presence of a light box behind the sample cell illuminates it from behind when taking images at $5$-second intervals using an SLR camera (D$90$, Nikon and Canon-SLR, Canon) that is automatically pre-set by a corresponding computer software.

\begin{table}
\caption{Properties of the water/glycerol solutions at $T$ = $25^{\circ}$C used in calculations.}
\resizebox{6cm}{!}{
\begin{tabular}{c c c c c c}
	\hline
	glycerol & density & viscosity & surface tension \\
	fraction & $\rho$  & $\mu$  & $\sigma$ \\
	$\%$ & (kg/m$^{3}$) &  (mPa$\cdot$s) &  (mN/m)\\
	\hline
	$0$ & $1000$ & $1.0$ & $72$ \\
	$40$ & $1117.5$ & $4.1$ & $65$ \\
	$50$ & $1150.6$ & $8.0$ & $64$ \\
	\hline
\end{tabular}
}
\end{table}

To further probe the kinetics of the infiltration process, we modify the viscosity of the primary fluid (water) by adding a concentration of glycerol (Sigma Aldrich) to create a water/glycerol solution. The addition of glycerol increases the viscosity of the fluid. The properties of the water/glycerol solutions are presented in Table 1. The density and viscosity values are calculated from Cheng $et$ $al$~\cite{Cheng00}. The densities of the water/glycerol solutions do not significantly change whereas the viscosities are increased by a factor up to $8$ times. The interfacial tension values are obtained from a study of interfacial tensions as a function of the volume of glycerol fraction performed by Shchekotov~\cite{Shchekotov10}. The interfacial tension values also do not significantly change. In addition, it has been determined from previous studies~\cite{Drelich96} that glycerol is hydrophilic and thus the contact angle of the water/glycerol solutions is essentially hydrophilic and is similar to water. 

For some infiltration experiments of more viscous fluids, we use a camera (Phantom) with a frame rate of $24$~fps to capture the infiltration and instability formation in slightly greater detail.

Before proceeding on how the infiltration phenomenon is influenced by physical parameters of the model soil, we probe the effects of the droplet impinging speed, $U_T$, on our system and we do not see any significant effect ~\cite{Wei13}. For the experiments described in this paper, we keep the droplet impinging speed constant at $U_T$ = $1.0$~m/s but vary the rain rat, $Q$, accordingly. 


\begin{figure*}
\includegraphics[width=6in]{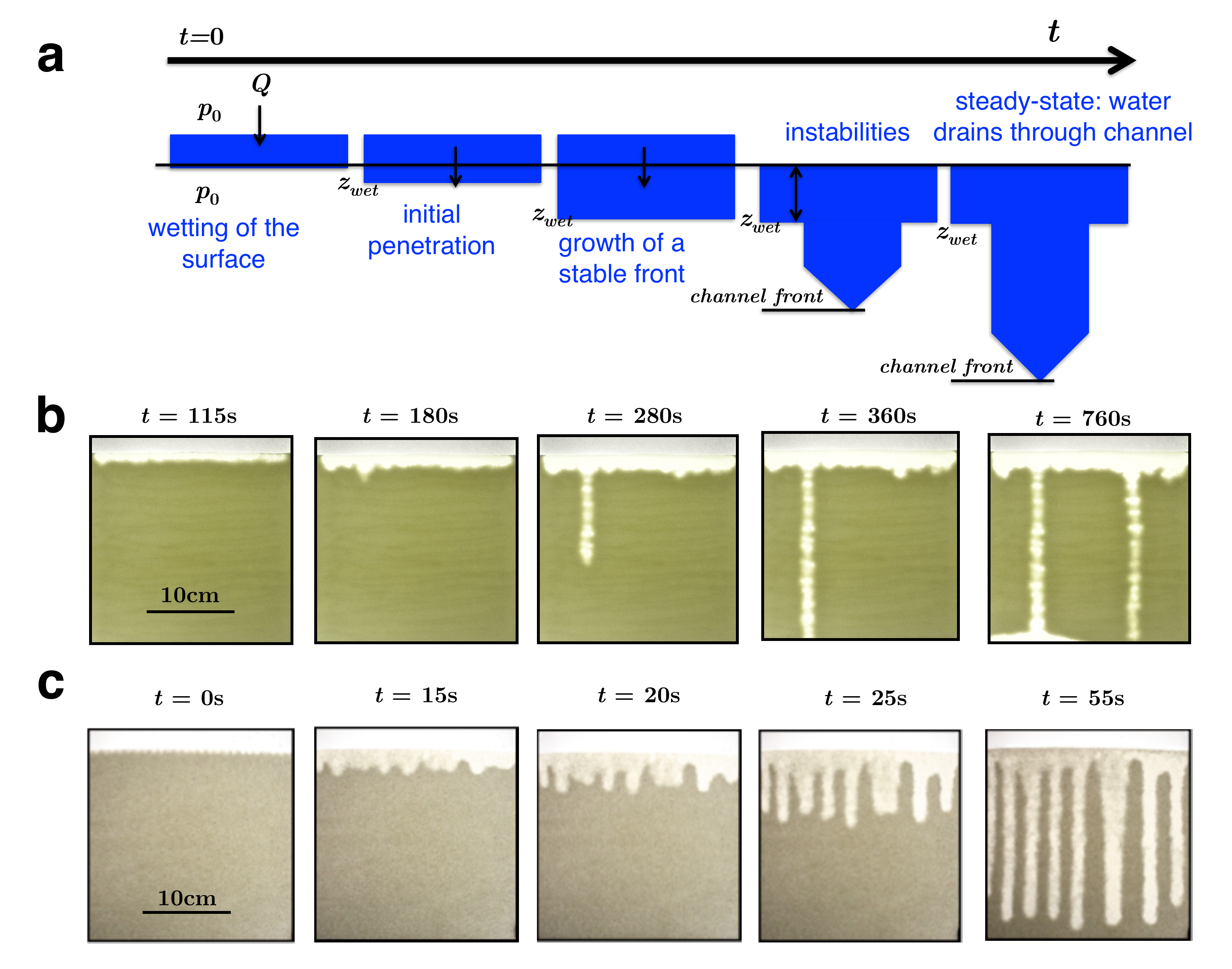}
\caption{(Color online) Sequence of images showing formation of water channels under steady rain. (a) Illustration of the formation of water channels in initially dry granular medium. This is preceded by a slow development of a homogeneous wetting front and growth of instabilities. The water channel serves as a preferential path for water drainage and is defined by a certain width, $d$. (b) Experimental images on infiltration of water channels for particle size $D$ = $300~{\mu}$m, viscosity $\mu = 1~$mPa$\cdot$s, and rain rate $Q=14.5~$cm/hr. The first water channel appears just after $t$ = $180$s after the formation of a homogeneous wetting front and only two water channels are formed. (c) Experimental images on the infiltration of water/glycerol solution in $D=1~$mm, $\mu = $4~mPa$\cdot$s, and $Q = 96.0~$cm/hr. At higher viscosities and higher flow rates, multiple water channels appear quickly and simultaneously.  }
\label{Fig2}
\end{figure*}



\section{Experimental Results}
\subsection{Experimental Observations - Infiltration Process}

In our experiments, infiltration under steady rain proceeds with rainwater initially wetting the soil surface as shown in Fig. $2$a. 

As rain is continuously supplied, a homogeneous wetting front begins to penetrate and develop inside the porous medium. As water continues to infiltrate vertically in the direction of gravity, the interface of the front eventually becomes unstable. Some of these instabilities fully develop into water channels while others do not. As soon as the water channels form, the wetting front ceases to infiltrate further deeper into the medium and these water channels serve as preferential paths for the drainage of water.  We study the infiltration process in initially dry and hydrophilic granular beads. First, we keep rain rate $Q$ constant but vary the diameter of the glass beads, $D$ = $2R$, which is proportional to the characteristic size of the pore~\cite{Culligan05}. Fig. $2$b shows a representative experimental image sequence for infiltration of water at $\mu$ = $1$~mPa$\cdot$s, $Q$ = $14.5$~cm/hr, and at $D$ = $300~{\mu}$m. In Fig. $2$b, we also observe a second finger instability, which results to a second water channel. 

Next, we keep the diameter $D$ constant but vary viscosity $\mu$ and rain rate $Q$. Fig. $2$c shows another representative image sequence for infiltration of water/glycerol mixture at $\mu$ = $4$~mPa$\cdot$s, $Q$ = $96.0$~cm/hr, and at $D$ = $1~$mm. 

In all these experiments, we measure the extent of the physical observations, such as the maximum width of the homogenous wetting front, $z_{wet}$, the average width or diameter of the channels, $d$, and the distance between two channels, $d^{\prime}$. These results concerning morphology are discussed in our companion paper~\cite{Wei13}. 

\subsection{Experimental Observations - Kinetics}

We can observe features common to all the performed experiments regardless of viscosity, flow rate, impinging speed, or bead diameters. Once rain begins to reach the soil, the first thing we observe is the establishment of a homogeneous wet front or wet zone, Fig. $3$. The development of the wet front may be slow or fast depending on the both properties of the fluid and the granular medium. The front gradually increases in size due to the presence of a continuous rain source, which supplies water to the top of the model soil surface. When it sufficiently forms, results suggest at first glance that the front is completely saturated. However, it has been shown that in the direction of gravity, a gradient of water saturation actually exists between the surface of the model soil and the area immediately just below it. This is to say that saturation levels are higher in the bottom of the front than at the top~\cite{Rezanezhad07, Cueto09}. Recent studies have suggested the role of saturation levels in the wetting front on instability formation although currently in literature, the wetting front is still a subject of ongoing investigations.

From experiments, the formation of the wetting front is particularly recognizable for fluids of low viscosities ($1$~mPa$\cdot$s). At larger viscosities, the homogeneous wetting front is easier to identify at higher flow rates. At larger viscosities but at lower flow rates, the wetting front appears faint because experiments give an impression that the viscous droplets do not spread enough to sufficiently coalesce with neighboring droplets.

We plot the average velocity of the wetting front, $v_{wet}$, as a function of bead diameter, $D$, in Fig. $4$a. The plot shows that the wetting front propagates at higher velocities in larger bead diameters than in smaller ones.  

\begin{figure}
\includegraphics[width=3.5in]{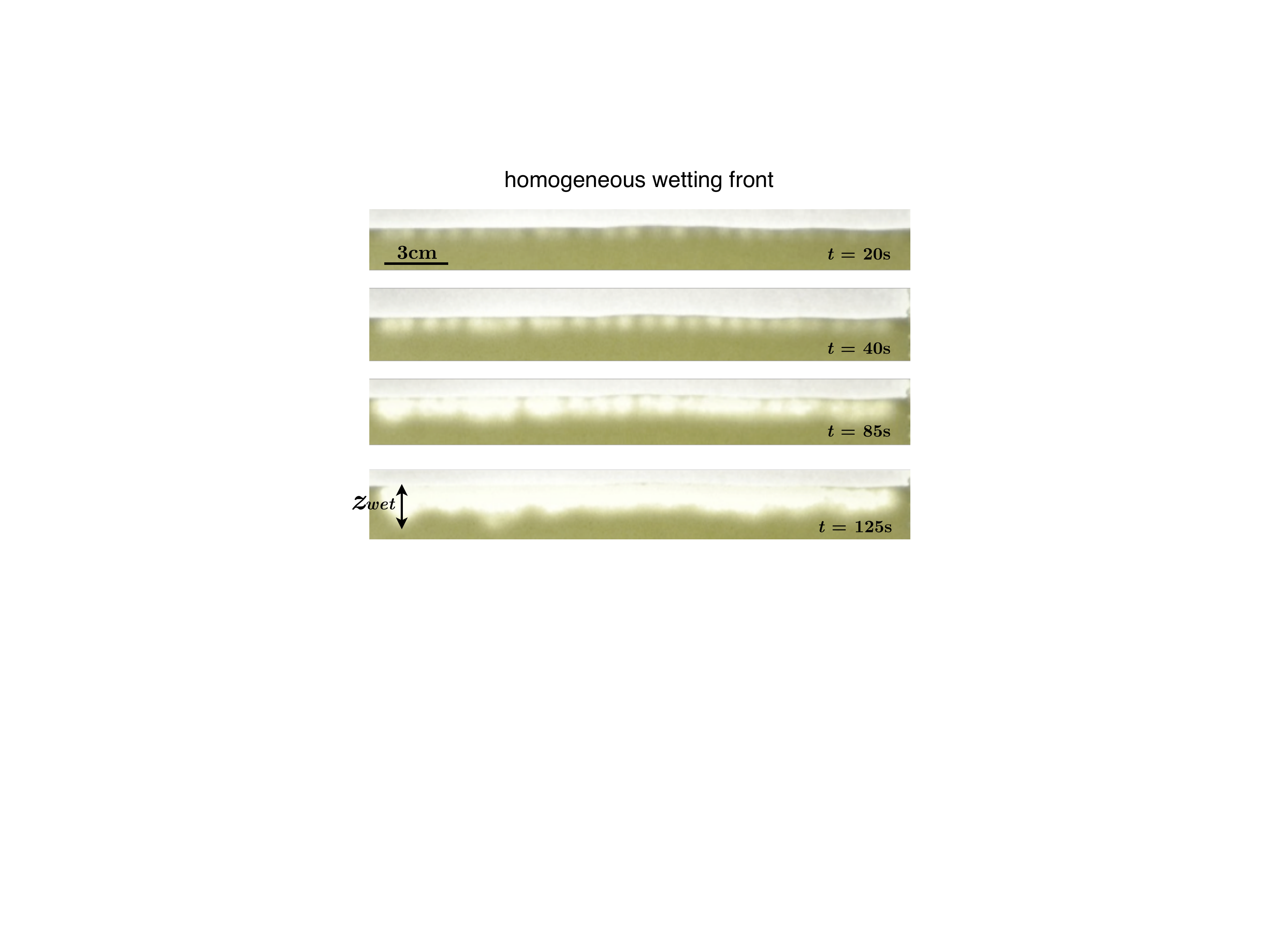}
\caption{(Color online). (a) Experimental images showing the evolution of the homogeneous wet zone with time during the early moments of rainfall. This is for the following experiment: droplet impinging speed, $U_T = 1.0$~m/s, particle size $D=300~\mu$m, viscosity $\mu = 1$~mPa$\cdot$s, and rain rate $Q=14.5$~cm/hr. }
 \label{Fig3}
\end{figure}

We also observe that the formation of the homogeneous wetting front takes time, particularly for low flow rates. In the case of low rainfall rate, this time scale can be measured by simple image analysis. Rain falling unto the soil surface allows the wetting front to expand downward in a homogeneous fashion, while moving at a certain velocity, $v_{wet}$. Eventually at a certain time, $t_C$, an instability occurs at the interface with the development of a finger. Fingers can appear either successively or simultaneously depending on the experimental conditions. The parameter $t_C$ refers to the time of formation of the channels and is the second main kinetics observation. This parameter will be discussed further in the next section. We will mainly focus on the time it takes for the first set of fingers to appear though data will be presented for the succeeding fingers.

Finally, once a finger is formed, the finger propagates deep in the soil until it reaches the bottom of the cell, where water drains. The propagation of the finger exhibits a certain velocity, $v$, making it the third kinetics observation. The plot of the velocities of the water channels as a function of time is seen in Fig. $4$b (at constant $\mu$, $U_T$, $Q$, but varying $D$) and Fig. $4$c (at constant $U_T$, $D$ but varying $\mu$, $Q$). In Fig. $4$b, we compare the finger channel velocities of both the first and second channel that develop. Results suggest that the velocities of the first channel are always greater than the velocities of the second channel.

Fig. $4$c shows the average velocities of the channel fronts as a function of rainfall flow rate, $Q$, and fluid viscosity, $\mu$. These data are taken using hydrophilic beads at constant bead diameter, $D$ = $1$~mm, and at constant $U_T$. In this figure, there is a clear trend pertaining to channel front (or finger) velocities with respect to rainfall flow rate and fluid viscosity. First, at constant fluid viscosity, $\mu$, the channel finger velocities increase with flow rate. The larger volume of fluid entering the soil results to faster propagation of channels downward. Second, at constant flow rate, $Q$, channel fingers in less viscous fluids ($1$~mPa$\cdot$s) propagate faster than in more viscous fluids. 

\begin{figure}
\includegraphics[width=3in]{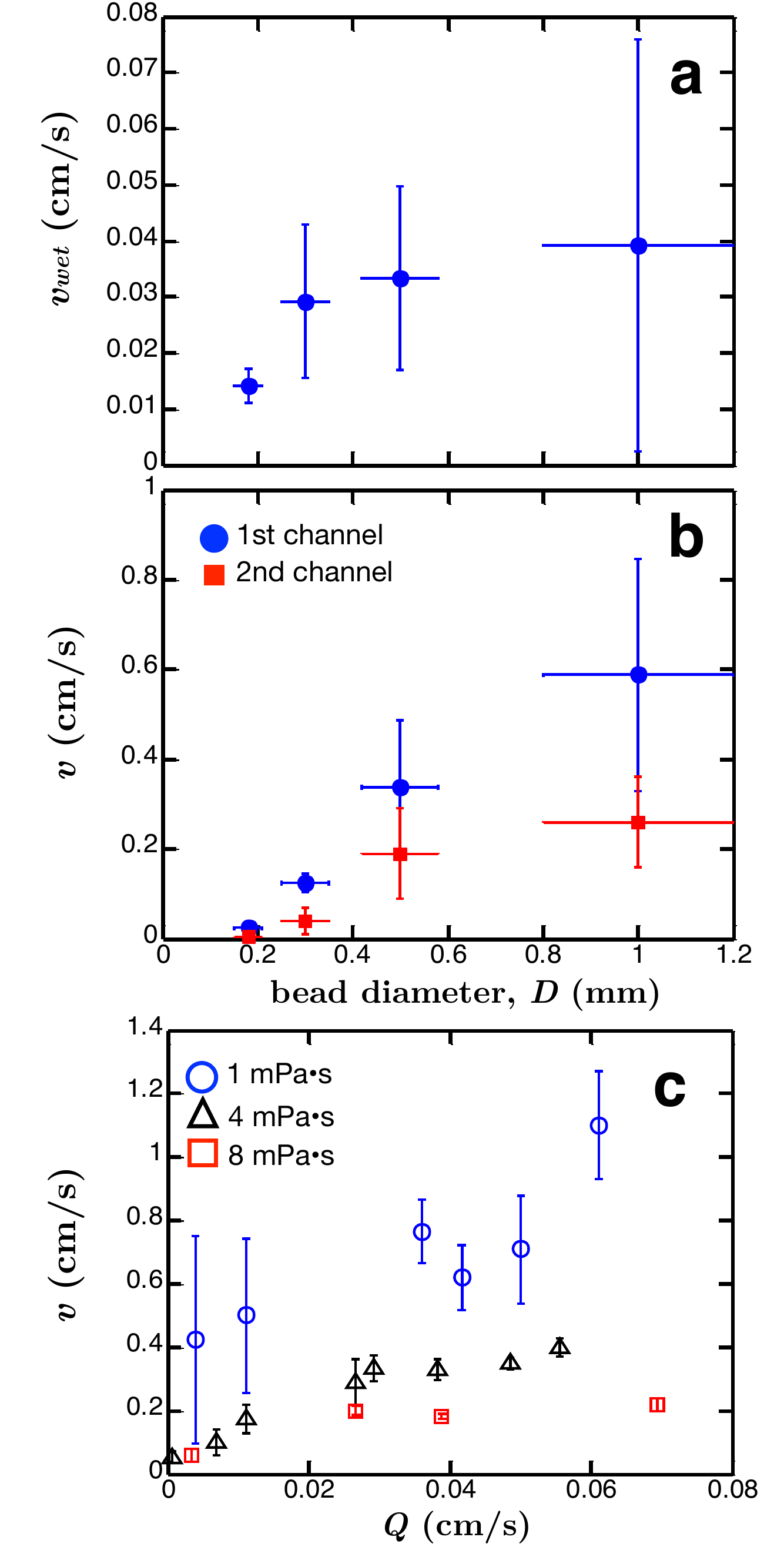}
\caption{(Color online). (a) Plot of the velocity of the wetting front, $v_{wet}$ as a function of bead diameter, $D$, at constant $U_T$, $Q$, and $\mu$. The graph shows that water moves faster in larger pores than in smaller ones. (b) Average channel front velocity of the first and second channels as function of bead diameter, $D$, at constant $U_T$, $Q$, and $\mu$. (c) Experimental data on the average channel front velocity, $v$, as a function of flow rate $Q$ also for different fluid viscosities, $\mu$, at constant $D$ and $U_T$.}
 \label{Fig4}
\end{figure}

Now that we have presented a general description of the experimental observations, we  look into them more closely in the next section. Among these observations, we first look into the onset of finger instabilities when the wetting front has fully formed. These instabilities serve as precursors for the formation of fluid channels.

\subsection{Time of formation of water channels}

In the infiltration process, the onset of the instability in the homogeneous wetting front corresponds to the transition towards  channel formation. Capillary forces dominate the formation of the wetting front. As the front develops inside the medium, capillary forces stabilize the interface while gravity has a destabilizing effect. At the onset of the instability, certain areas of the homogeneous front develop relatively faster than others. Many of these proto-fingers ~\cite{Cho02} develop in the front but only one or a few mature and grow into a full water channel. Once a finger fully grows into a channel, the other proto-fingers cease to develop.

During the growth of the instability, the entire wetting front still continues to move in the direction of gravity as water is still continuously supplied at the surface. When the finger instability grows into a water channel, only then will the front plane stop growing. This is because the water channel serves as a preferential path for water drainage as it provides an outlet for water. 

Expectedly for different bead sizes, the time of appearance of the water channel also varies as water flow through a porous medium is limited by the size of the pore. It takes a longer time for water to flow through smaller pores than it takes through larger pores. Thus, it takes a longer time before the water channel appears. Fig. $5$a shows different time scales observed in our experiments. In this graph, we consequently plot the time of appearance of the first and second channel, $t_{C_1}$ and $t_{C_2}$, respectively. Using water ($\mu$ = $1$~mPa$\cdot$s) and also at constant $Q$, $U_T$, we consistently observe two channels regardless of $D$. There is also clearly a trend for the formation of the first water channel as a function of particle size. However, the second channel does not exhibit such a clear trend. We also plot in the same graph the quantity $t_h$, which is the characteristic time when the front becomes homogeneous. This is due to the experimental design, in which the front during the first few seconds of rain is not homogeneous as previously shown in Fig. $3$.

In the infiltration of more viscous fluids using constant $D$ and $U_T$ but at varying $Q$, the time of appearance of the formation of water channels seems to generally decrease with increasing flow rate as shown in Fig. $5$b. This physically means that as more volume of water enters the medium at high flow rates, water immediately requires a drainage outlet and thus channels form rather quickly. The rightmost point in this graph, however, corresponds to a more viscous fluid ($8$~mPa$\cdot$s) infiltrating at extremely high flow rate. Experiments show that instead of forming distinct water channels having widths considerably less than the length of the $2$D cell, a massive front is generated that covers the entire length of the cell.

With these observations, we use a proposed model from literature to explain the physics of the phenomenon. 

\begin{figure}
\includegraphics[width=3.3in]{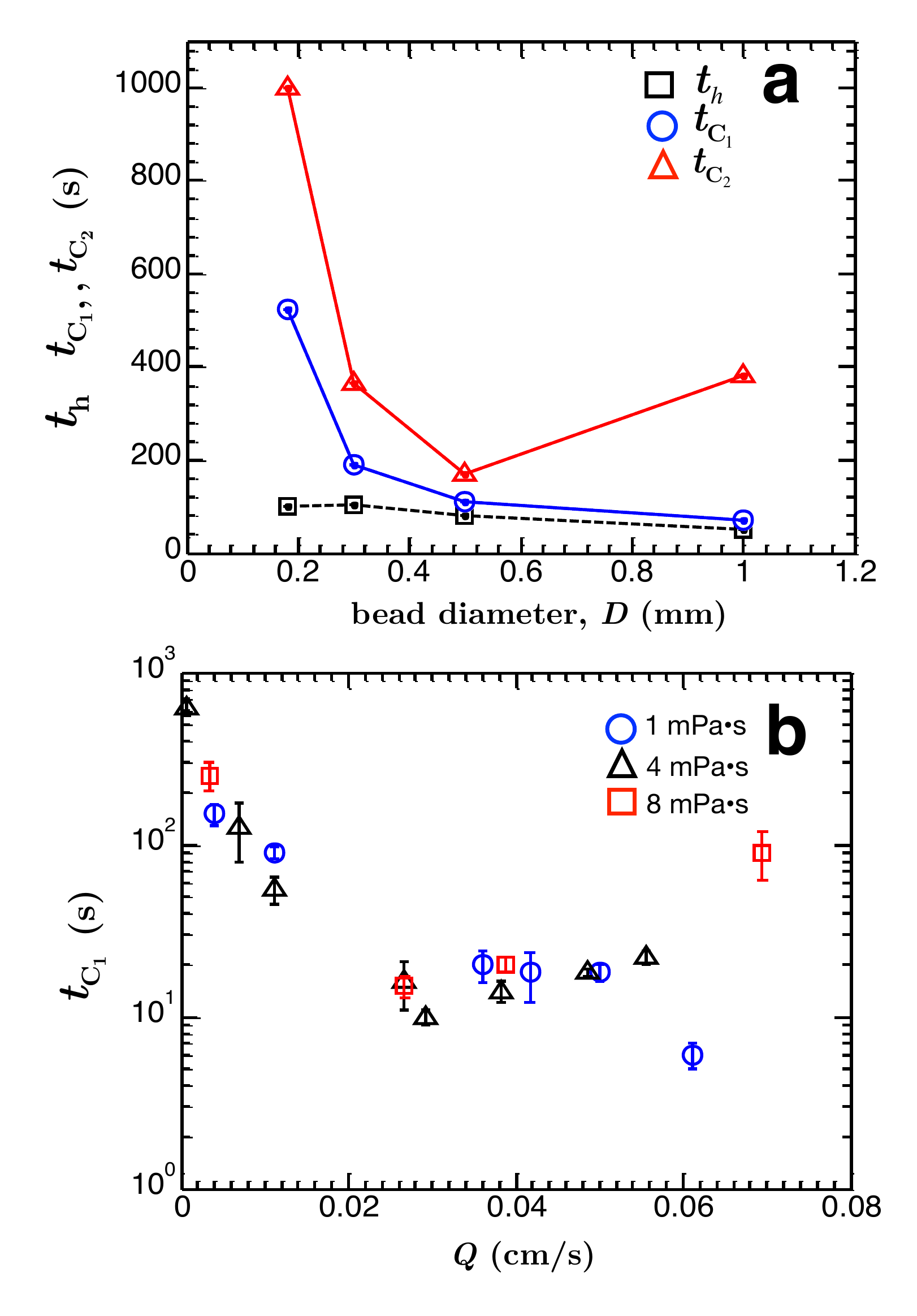}
\caption{(Color online).  (a) Plot of the time of formation of the first channel, $t_{C_1}$, and second channel, $t_{C_2}$, as well as the time it takes for the front to become homogeneous, $t_h$, all as a function of bead size diameter, $D$, at constant $Q$, $\mu$, and $U_T$. A clear trend exists during the formation of the first channel, meaning water channels form much later in smaller particles where liquid flow is much slower. However, the formation of the second channel seems to be conditioned not just by particle size but also by other parameters, which are yet to be fully determined. A third quantity $t_h$ is the time when the front becomes homogeneous. Owing to the unique design of the rain source, it takes time for neighboring droplet impact sites to coalesce and form a continuous front. (b) Experimental data of average time of formation of the first water channel, $t_{C_1}$, as a function of flow rate $Q$ for different fluid viscosities, $\mu$, but at constant particle size diameter $D$ and droplet impinging speed, $U_T$. There appears to be a decreasing trend with respect to flow rate $Q$ initially at low flow rates but  this trend slowly increases at higher flow rates and higher viscosities.  }
 \label{Fig5}
\end{figure}


\section{Model Discussion}

The linear stability approach has been used in numerous studies. Through experimental results, it became more apparent that infiltration is a form of immiscible fluid displacement between a wetting phase (liquid) and a non-wetting phase (air). To study such a phenomenon, Saffman and Taylor~\cite{Saffman58} performed one of the pioneering approaches on the subject using experiments in Hele-Shaw cells filled with two fluids of different viscosities using the fact that flows in porous media and in Hele-Shaw cells are formally analogous. The different properties of both fluids result to perturbations occurring in the interface. These perturbations develop into instabilities. Crucial to the analysis of the formation of the instability is the definition of the pressure at the interface of these two fluids. Saffman and Taylor~\cite{Saffman58} notes that a sharp interface is nonexistent but nevertheless assumed that there is no pressure jump across the interface since the characteristic width of the perturbations in the interface is smaller than the length scale of the motion. Thus, for Saffman and Taylor~\cite{Saffman58}, pressure is continuous. This results to an equation where any perturbation, whether large or small, can grow into a water channel. This contradicts our observation where the finger size is clearly defined from a characteristic perturbation that develops the fastest. Chuoke $et$ $al$~\cite{Chuoke59} incorporated this limitation in the modification of the original analysis of Saffman and Taylor. In their assessment, the pressure is in fact not continuous and the discontinuities are defined by a Young-Laplace relationship. Hence, the interfacial pressure jump was described by an effective macroscopic surface tension. However, it is often difficult to determine exactly the effective macroscopic surface tension and thus Parlange and Hill~\cite{Parlange76} later argues that this might only be valid for fluid displacements in parallel plates and not for porous media such as soil. The analysis of Saffman and Taylor~\cite{Saffman58} and Chuoke $et$ $al$~\cite{Chuoke59} were performed for a less viscous fluid driving a more viscous one. While the opposite of the infiltration process described in this paper where the more viscous fluid, water, displaces the less viscous fluid, air, its principles are certainly analogous to water infiltration in a dry porous medium. Nevertheless, Parlange and Hill~\cite{Parlange76} further proposed another approach for calculating channel width taking into account the influence of soil-water diffusivity when the curved front propagates. While using the same basic principles of linear stability analysis initially described by Saffman and Taylor~\cite{Saffman58}, Parlange and Hill~\cite{Parlange76} considered the front as a discontinuity and assumed that if $u_1$ is the velocity of the relatively flatter front, then the velocity of the curved front, $u_2$, is decelerated proportionally to its curvature, ($r_1^{-1}$ + $r_2^{-1}$) according to Eq. 1, where $r_1$ and $r_2$ are the front$^{\prime}$s two principal radii of curvature:

\begin{equation}
   u_2 = u_1 - \xi \left( \frac{1}{r_1}+\frac{1}{r_2} \right),
\label{Eq1}
\end{equation}
where $\xi$ is a function describing soil properties. In other words, Parlange and Hill~\cite{Parlange76} described the interfacial pressure as a function of front velocity.

Assuming that the fluid is incompressible and the porosity of the granular material is uniform, the velocity potential satisfies Laplace$^{\prime}$s equation, ${\nabla^2}{\phi}$ . Darcy$^{\prime}$s law is then used to describe the velocity of the front in the $z$ direction, where $z$ is pointing downward:

\begin{equation}
   q_z =  - \frac{\kappa_s}{ \left(S_s - S_0 \right)} \nabla \phi,
\label{Eq2}
\end{equation}
where $\kappa_s$ is the hydraulic conductivity, $S_s$ is the saturated water content, and $S_0$ is the initial water content. The hydraulic conductivity $\kappa_s$ measures the ease in which a fluid flows through pore spaces~\cite{Hillel03}. As $\kappa_s$ will appear in succeeding equations, it is worthwhile to note its definition~\cite{Verneuil11}.

\begin{equation}
   K = \frac{\rho g K_0 D^2}{\mu},
\label{Eq3}
\end{equation}
where $\rho$ is the fluid density (water), $\mu$ is the dynamic viscosity, $g$ is the acceleration due to gravity, $D$ is the particle diameter, and $K_0$ is the intrinsic permeability. For a random close packing of spheres having porosity, $\epsilon$ = $0.36 - 0.40$, $K_0$ can be determined using the Karman-Cozeny equation~\cite{Verneuil11}. From this approach, $K_0$ = $6.3$x$10^{-4}$. From Eq. 3, conductivity is proportional to the square of the particle size, so we expect that water infiltration proceeds extremely faster in larger bead sizes as indeed observed from our experiments.

From an initial condition of $z = 0$, taking the derivative of the front position with time results to the velocity of the curved interface (see Parlange and Hill~\cite{Parlange76} for more details on the linear stability analysis):
\begin{equation}
   u_2 =  u_1 + {a}{\lambda^2}{\xi} \exp \left({i}{\lambda}{y}+{\omega}{t}\right),
\label{Eq4}
\end{equation}
where $a$ is the amplitude, $\lambda$ is the wavelength, $\omega$ is the growth rate of the instability, and $\xi$ is a function describing soil properties defined as: 

\begin{equation}
   \xi =  \int_{S_0}^{S_s}{ \frac{D_f}{S_s - S_0}} \mathrm{d}\theta,
\label{Eq5}
\end{equation}
where $D_f$ is hydraulic diffusivity, which varies with water content, $S$, in this equation. The hydraulic diffusivity is defined as the ratio of the flux to the soil-water content gradient~\cite{Hillel03}. It is to note that water movement in soil is not actually described as diffusion, in the strictest sense, but of mass flow or convection~\cite{Hillel03}, although the term diffusivity has been used for historical reasons.

Nevertheless, the solution to the Laplace equation such that  $\omega$ $>$ $0$, gives: 

\begin{equation}
   \omega = \lambda \left( \frac{\kappa_s - u_1\left( S_s - S_0 \right) }{ S_s - S_0 } \right) - {\xi}{\lambda^2},
\label{Eq6}
\end{equation}
If pressure is continuous across the front, Parlange and Hill~\cite{Parlange76} notes that the instability that grows the fastest and results to a channel satisfies d$\omega$/d$\lambda = 0$, where $\lambda$ is given by:

\begin{equation}
   \lambda = {\frac{1}{2}}\left( \frac{\kappa_s - u_1(S_s - S_0)}{\int_{S_0}^{S_s}{D_f}\mathrm{d}\theta}\right),
\label{Eq7}
\end{equation}

Using substitution of Eq. 7 to Eq. 6, we obtain the growth rate of the unstable wavelength, $\omega$ = $\xi$$\lambda^2$, where $\lambda$ is related to the equation for determining finger width or diameter, $d$. Parlange and Hill~\cite{Parlange76} argues that the finger width is roughly of this dimension, $d$ = $\pi$/$\lambda$, and the soil diffusivity can be expressed in terms of soil sorptivity, written as:

\begin{equation}
   s_w^2 = {2\left(S_s - S_0\right)}{\int_{S_0}^{S_s}{D_f}\mathrm{d}\theta}
\label{Eq8}
\end{equation}

Sorptivity is the measure of the capacity of a medium to absorb or desorb liquid through capillary forces~\cite{Philip69}. Culligan $et$ $al$~\cite{Culligan05} states that the sorptivity depends on the properties of both the fluid and the porous material. Using scaling analysis, Culligan $et$ $al$~\cite{Culligan05} used experiments in real sandy soil to arrive at the following relationship for sorptivity:

\begin{equation}
   s_w = {s^*} \left( {\frac{\epsilon l^* \sigma cos\theta^*}{\mu S_{av}^{c - 1}}}\right)^{1/2},
\label{Eq9}
\end{equation}
where $s^*$ is the dimensionless intrinsic sorptivity with a value equivalent to $s^* = 0.133$ as experimentally determined for sandy-type soil~\cite{Culligan05}, $\epsilon$ is the porosity, $\rho$ is the fluid density, $\mu$ is the dynamic viscosity, $l^*$ is a microscopic characteristic length scale of the medium, $\sigma$ is the surface tension, $\theta^*$ is the effective contact angle, $S_{av}$ is the average saturation of the infiltrating fluid at the inlet of the porous medium, and $c$ is an empirical coefficient determined from the Brooks-Corey pore size distribution index. We can expect $l^*$ to be proportional with the particle size diameter, $D$. However, it is difficult to determine the value of $S_{av}^{c - 1}$; thus, we have assigned $\frac{l^*}{S_{av}^{c - 1}}$ in Eq. 9 to be equivalent to $\beta D$, where $\beta$ is a fitting parameter equal to $\beta$ = $0.015\pm0.002$. The value of $\beta$ is kept constant for all equations where this parameter appears. Thus, Eq. 9 is now simplified into the following equation:

\begin{equation}
   s_w = {s^*} \left( \frac{\epsilon \beta D \sigma cos\theta^*}{\mu}\right)^{1/2}.
\label{Eq10}
\end{equation}

Further substitutions result to the equation for determining channel width, which is written as:

\begin{equation}
   d =  \frac{\pi s_w^2}{\kappa_s \left(S_s-S_0\right) \left( 1- Q/\kappa_s \right) } .
\label{Eq11}
\end{equation}

The parameter $S$ is the water content defined as the ratio of the volume of water in the soil and the total volume of the soil. The subscripts $s$ and $0$ respectively represent the saturated state and initial state of the soil. Since these experiments have been performed from an initially dry and random close-packed glass beads, $S_0$ = $0$ and $S_s$ = $\epsilon$, where $\epsilon$ is the porosity, representing the maximum amount of water that can be contained within the pore spaces. Eq. 11 has shown good agreement with experimental results obtained from sand~\cite{Glass91, Glass89, Selker92}. 

The measure of the capillary forces is manifested in the surface tension factor in the sorptivity. As capillary forces are increased, so does the sorptivity. This consequently leads to an increase in channel width or size. Capillary forces stabilize the wetting front. As the instability develops, the characteristic size of the perturbations that can develop also increases with increasing capillary forces. This is the fundamental reason why soil with smaller bead diameters shows larger channel widths at constant viscosity and flow rate. 

We can also use the same analyses as a starting point to calculate the characteristic time it takes for a wavelength to become unstable. The linear stability analysis also provides the period, $\tau$, which is given by $\tau$ = $2\pi$/$\omega$, where $\omega$ = $\xi$$\lambda^2$. We can calculate for $\lambda$ via substitution of Eq. 5 to Eq. 8 to yield the following relationship:

\begin{equation}
   \xi = \frac{s_w^2}{{2\left(S_s-S_0\right)}^2}.
\label{Eq12}
\end{equation}

Furthermore, $\tau$ can be further simplified into:

\begin{equation}
   \tau = \frac{4 \epsilon^2 d^2}{s_w^2 \pi},
\label{Eq13}
\end{equation}
where ($S_s$ - $S_0$) $\approx$ $\epsilon$ and $\epsilon$ is porosity, $d$ is the channel finger width (Eq. 11) and $s_w$ is sorptivity (Eq. 10). This time scale reflects the time of appearance of the first channel that develops from the instability of the homogeneous wetting front. We put focus on the formation of the first water channel because experimentally a clear trend is observed with respect to bead size as shown in Fig. $5$a. Combining Eq. 3, 10, 11, and 13 gives the following scaling for $\tau$:

\begin{equation}
   \tau \sim \frac{\mu}{D^3 \left(1-\frac{\mu Q}{\rho g K_0 D^2}\right)^2} ,
\label{Eq14}
\end{equation}

Parlange and Hill$^{\prime}$s main contribution has been to describe the beginning of the instability and the morphology of the fingers that develop from such an instability. The model is the basis of subsequent analysis from Glass $et$ $al$~\cite{Glass89} in describing the channel finger propagation velocity, $v$, to arrive at the following relationship:

\begin{equation}
   v = \frac{\kappa_s}{\left(S_s-S_0\right)}f(Q/\kappa_s).
\label{Eq15}
\end{equation}

Further analysis by Glass $et$ $al$~\cite{Glass89} and Wang $et$ $al$~\cite{Wang98} shows that Eq. 15 can be written as follows:

\begin{equation}
   v = \frac{\kappa_s}{\epsilon} \left(C + (1 - C) \sqrt{Q/\kappa_s}\right),
\label{Eq16}
\end{equation}
where $C$ is the projected zero flow velocity for fingers~\cite{Wang98}, which is dependent on the dimensionality (whether $2$D or $3$D systems) of the granular system. 

Further expansion of Eq. 16, using substitution of the definition of hydraulic conductivity (Eq. 3) yields the following scaling:

\begin{equation}
   \mu v \sim (\mu Q)^{1/2}.
\label{Eq17}
\end{equation}

In Eq. 17, the dependence of the channel finger velocity, $v$, on $\sqrt{Q}$ is demonstrated for all fluid viscosities. 

We have so far discussed the evolution of the instability formation. Instability precedes the formation of the water channel and we have shown that the drainage of water channels from an initial wetting front can be described by linear stability analysis. More recently, numerical simulations performed by Cueto-Felgueroso and Juanes~\cite{Cueto-Felgueroso08, Cueto-Felgueroso09} have also advanced our understanding of the topic. The analysis of Cueto-Felgueroso and Juanes~\cite{Cueto-Felgueroso08, Cueto-Felgueroso09} proposes a macroscopic phase-field model during unsaturated flow. They also employed linear stability analysis to stress the importance of the role of the water saturation in the growth of the instabilities. In essence, their model introduces a non-linear term to the classical Richards equation to account for the appearance of perturbations. This term is formally related to the surface tension at the interface of the wetting front. From mathematical calculations, they predict that finger width and finger velocity both increase with infiltration rate. 

Nevertheless, despite many proposed modifications to existing models, linear stability analysis is enough to describe well the morphology and certain aspects of kinetics of the formation of water channels as will be discussed in the next section.


\section{Model Application and Comparison with Experiments}

Owing to our quasi-$2$D rainfall set-up built with equally spaced capillary tubes, it takes a certain time $t_h$  for the front to establish homogeneously.  We take into account the previously calculated parameter, $t_h$, as a delay during which the front becomes fully homogeneous. The effective time of appearance of the first channel is then given by:

\begin{equation}
 t_{C_1} = t_h + \tau,
\label{Eq18}
\end{equation}
where $\tau$ is calculated from Eq. 14 and $t_h$ still remains to be estimated.

\subsection{Time scale for the establishment of the homogeneous zone}

We can compute for $t_h$ since it is mainly a function of the distance between the capillaries, $d_{cap}$, in the rain set-up and flow rate, $Q$, as schematically shown in Fig. 6a. 

We let $Q$ be the total flow rate impacting the system. The raindrops will impact on the surface of certain initial volume. Successive impacts will increase the volume of the drop, which will eventually result to coalescence of neighboring droplet impact sites. The total volume then on one impact site underneath a capillary tube is a function of time, $V(t)$ = ${A}{Q}{t}$, where $A$ is the cross-sectional area of the cell (m$^2$). Note that $Q$ has units in velocity (m/s). When the front makes a depth in the medium equal to $d_{cap}$, the time it takes for the front to become sufficiently homogeneous can be calculated from the parameters of the sample cell, where:

\begin{equation}
 t_h = \frac{\epsilon d_{cap}}{Q} ,
\label{Eq19}
\end{equation}
where the spacing between capillary tubes in the rain set-up is set at $d_{cap}$ = $1$~cm, $Q$ is the total flow rate in units of velocity, and $\epsilon$ is the porosity of the medium. The size of the sample cell is also a factor but $l$ x $e$, where $l$ and $e$ are the length and thickness of the sample cell respectively, will just cancel out with the cross-section $A$ in the denominator. Thus, in this equation, the spacing of the capillary tubes in the rain source is an important criterion that influences the front homogenization. This means that if the spacing had been larger, e.g.  $d_{cap} > 1$~cm, we can expect the front homogenization to occur much later due to the fact that coalescence of droplet impact sites is less favored when the distance between them increases. 

Once the front becomes homogeneous, it propagates in a uniform manner downward until instabilities develop. 


\begin{figure*}
\includegraphics[width=6in]{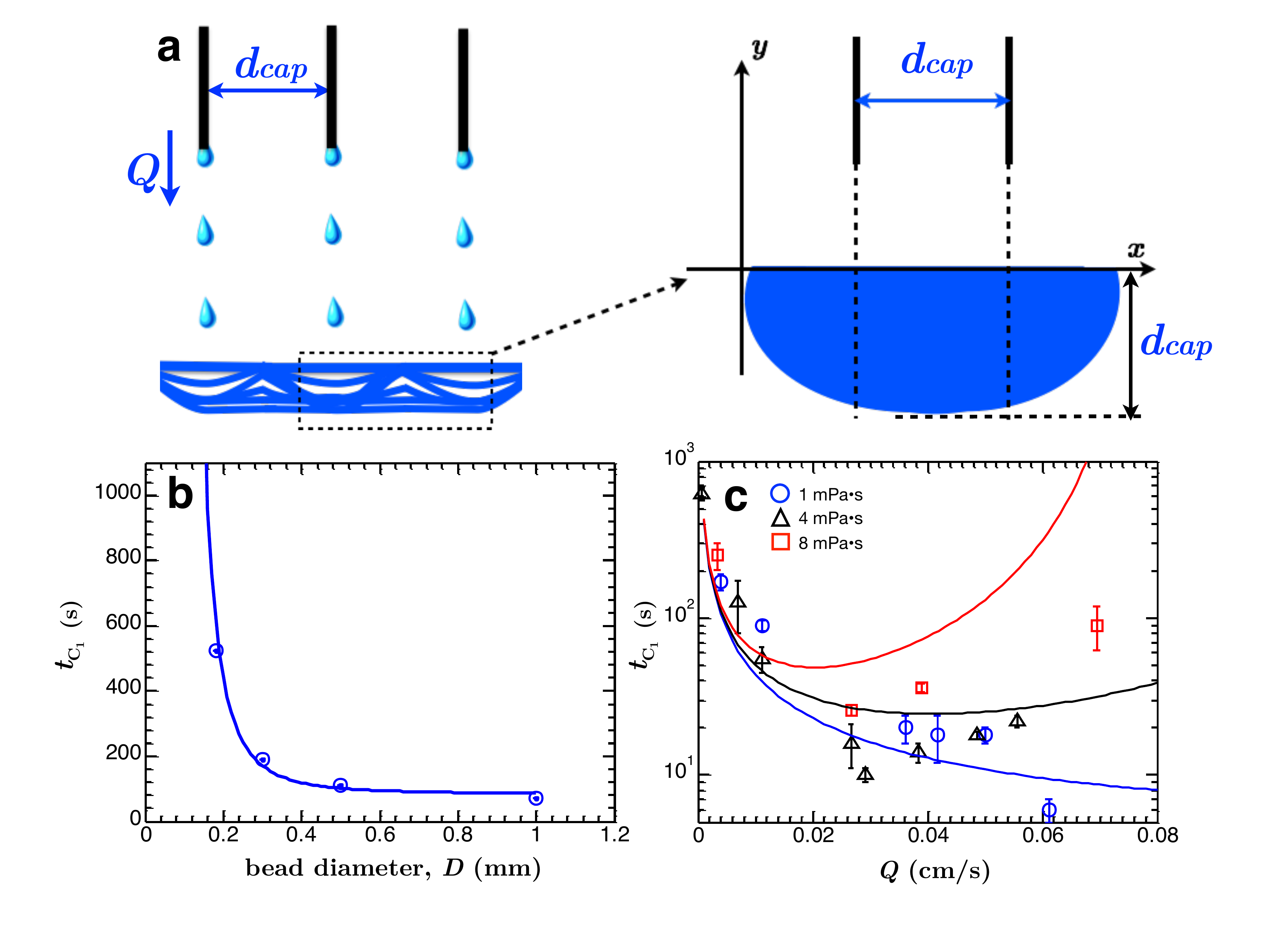}
\caption{(Color online). (a) Illustration of the first few moments of rainfall. Due to experimental design, there is a time scale, $t_h$, at which the front becomes homogeneous. As the droplets impact the granular medium of an initial volume, successive drops will increase this volume as a function of time until neighboring impact sites coalesce to form a homogeneous wetting front. We suppose that this coalescence is achieved when the depth of the front is equal to the spacing between adjacent capillary tubes. (b) At constant $U_T$, $Q$, and $\mu$, experimental data on time of appearance and formation of the first water channels as function of bead size, $D$. The time of formation of the first water channel, $t_{C_1}$, fits well with a model (Eq.~\ref{Eq18}) derived from the linear stability analysis of a wetting front. (c) At constant $U_T$ and $D$, we plot the experimental data for different $\mu$ as a function of $Q$ and show considerable agreement with the model (Eq.~\ref{Eq18}), particularly at low flow rates. The model predicts an initial decrease in $t_{C_1}$ at low $Q$, but will gradually increase at higher $Q$, especially for higher values of $\mu$. }
 \label{Fig6}
\end{figure*}


\subsection{Time scale for the instability to develop}

At constant $U_T$, $Q$, and $\mu$ but at varying $D$, Eq. 18 agrees well with experiments as shown in Fig. 6b. In this figure, $t_{C_1}$ decreases with increasing $D$. In addition, as a function of flow rate $Q$ and for three different viscosity values, $\mu$, Eq. 18 also shows decent agreement with experiments as depicted in Fig. 6c. This figure shows interesting behaviors. At low viscosities (water),  $t_{C_1}$ decreases with $Q$, meaning water channels will form faster at higher flow rates. However, as the viscosity increases the time of formation of water channels initially decreases at low $Q$ but then slowly increases as $Q$ further increases. This becomes even more prominent at higher viscosity values (8~mPa$\cdot$s), where larger flow rates increases the time it takes for water channels to form. Based on experimental observations at high viscosity and high flow rate, where $Q$ is close to the value of $\kappa_s$, the fluid initially infiltrates as one massive front, so water channels form at a later time.

The decent agreement between our experimental data and the theoretical results suggest that taking into account an additional time delay for the formation of the homogeneous front is necessary to obtain a more accurate description of the process, especially at constant $U_T$, $Q$, and $\mu$ as shown in Fig. 6b. In Fig. 6c however, the model, which is based on the linear stability analysis developed by Parlange and Hill, seems only to capture the time scale for the destabilization of the homogeneous front for low flow rates. The model appears to be far less accurate when $Q$ approaches $\kappa_s$, which happens at conditions of higher $\mu$ and higher $Q$ values. It would then be worth testing these experimental data using other models in literature, in particular the recent model developed by Cueto-Felgueroso and Juanes~\cite{Cueto09} to check whether the predictions are better. This is the subject of future investigations.

\subsection{Channel finger velocity}

We can derive the velocities of the channel from Eq. 16. Similarly, we apply this equation to two different cases, first at constant $Q$, $\mu$, and $U_T$ but at varying $D$ and second at constant $D$ and $U_T$ but at varying $Q$ and $\mu$. The results of the first case are shown in Fig. 7a. This figure shows that the velocity of the water channel is also dependent on the particle size. Larger particles have larger pores and thus have greater water flow velocity, allowing water to easily flow down. 

The results of the second case, on the other hand, are shown in Fig. 7b. In both cases in applying Eq. 16, we use $C = 0.2$.

In Fig. 7a, higher channel finger velocities are predicted in larger bead diameters due to the accompanying larger pore size. In Fig. 7b, the dependence of the channel finger velocity, $v$, on $\sqrt{Q}$ is demonstrated for all fluid viscosities as predicted in Eq. 17.

From these results, the model derived from linear stability analysis fits reasonably well with the experimental data.


\begin{figure*}
\includegraphics[width=6in]{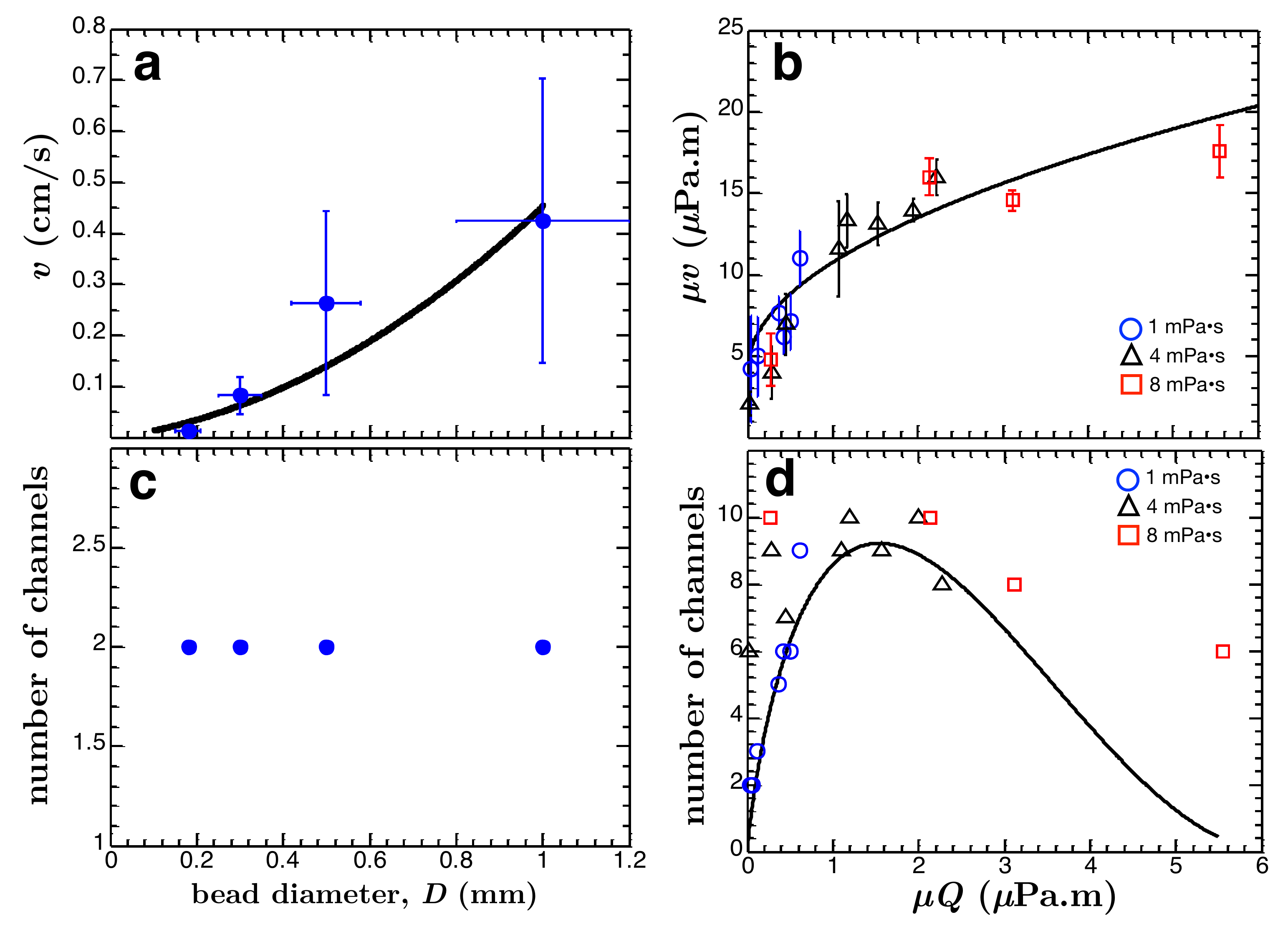}
\caption{(Color online). (a) Experimental data on the average channel front velocity, $v$, as a function of $D$ at constant $Q$, $\mu$, and $U_T$. Model fit shown by the solid line is Eq. 15. (b) Experimental data on the average channel velocity, $v$, rescaled with viscosity $\mu$, as a function of flow rate $Q$ also for different fluid viscosities using Eq. 16. The relationship between $\mu v$ and $\mu Q$ is further emphasized in Eq. 17, where the dependence of $(\mu v)$ on $\sqrt{\mu Q}$ is demonstrated. These are results from experiments performed at constant $D$ and $U_T$. (c) Experimental data of number of channels, $N$, as function of $D$ at constant $Q$, $\mu$, and $U_T$. (b) Experimental data of number of channels observed, $N$, as a function of flow rate $Q$ and for different fluid viscosities, $\mu$, rescaled using Eq. 21 at constant $D$ and $U_T$. }
 \label{Fig7}
\end{figure*}


\subsection{Number of channels}

Now that we have shown the model fits for channel width, characteristic time of channel formation, and channel finger velocity, in this section we apply the same model to predict the number of channels, $N$. At constant $U_T$, $Q$ and $\mu$ but varying $D$, we observe two channels plotted in Fig. 7c and the separation distance between them roughly remains constant as shown in our companion paper~\cite{Wei13}.

At constant $D$ and $U_T$ but at varying $Q$ and $\mu$, shown in Fig. 7d, the number of channels observed for less viscous fluids ($1$~mPa$\cdot$s) such as water generally increases with $Q$. The value of $N$ in more viscous fluids, however, decreases with increasing $Q$. At low $Q$, there are less number of channels in less viscous fluids but already more channels in more viscous fluids. This result already provides a clue that the fluid viscosity modifies the temporal dynamics of the instability formation. 

Similar to aforementioned approaches, we rescale the experimental data by using the equation for channel width obtained from linear stability analysis as a starting point. 

By mass conservation, the total volumetric flow rate, $QA$, is equal to the volumetric flow rate in each finger multiplied by the number of fingers, $N$. This relationship can be written as:

\begin{equation}
 Q A = N q_f,
\label{Eq20}
\end{equation}
where $Q$ is the total flux into the granular system (m/s), $A$ is the total cross-sectional area, $q_f$ is the flux through each finger, which is a function of channel velocity, $v$ (Eq. 15), and channel finger width, $d$ (Eq. 11). Further expansion of Eq. 20 approximately results to:

\begin{equation}
 N \sim \frac{A}{D^2}\left( \frac{\mu Q}{\rho g K_0 D^2} \right)^{\frac{1}{2}}\left(1-\frac{\mu Q}{\rho g K_0 D^2} \right)^2
\label{Eq21}
\end{equation}

In Eq. 21, we can see that $N$ exhibits two behaviors as a function of fluid viscosity and flow rate at constant bead diameter. It increases with $\sqrt{\mu Q}$  but decreases with $(1 -\frac{ \mu Q}{\rho g K_0 D^2})^2$ . The curve is presented in Fig. 7d.

In this figure, the number of channels initially increases at low viscosity fluids ($1$~mPa$\cdot$s) and low flow rates. However, at higher fluid viscosities ($8$~mPa$\cdot$s) and higher flow rates, $N$ reduces in value not because there are no individual channels that form but because the fluid eventually infiltrates as one massive stable front. Additional points are indeed necessary to ascertain the decrease of the curve especially within the range of $3$x$10^{-6}$ $<$ $\mu Q$ $<$ $6$x$10^{-6}$. It is however experimentally difficult due to the limitations of the size of the rain source  and the available capillary tubes.

At constant bead size, increasing the viscosity reduces the hydraulic conductivity of that particular fluid; thus, the flow rate, $Q$, slowly approaches the value of $\kappa_s$ ($Q \rightarrow \kappa_s$). When this happens, the number of channels decreases as predicted by the equation. Physically this implies that for a certain total cross-sectional area of the cell, we can predict the number of channels that can appear during the infiltration of a fluid within that area. And that there is a maximum number of channels that can form within the limits of the cross-sectional area as a function of rainfall flow rate and fluid viscosity. For example, if rain impacts a cross-sectional area of $20$~cm$^2$, then we will obtain the maximum number of channels when $\mu Q \approx 1.5$x$10^{-6}$~$\mu$Pa$\cdot$m. This means the maximum is achieved either using low viscosity fluids but infiltrating at high flow rates or using higher viscosity fluids infiltrating at lower flow rates.


\section{Conclusion}

Preferential water paths are drainage outlets. Once they form, they effectively reduce the water content around the root zone. Studies have shown how this is affected by properties of the granular material, such as pore size. But through extensive experimental results, we have also explored the influence of the fluid properties as well on the formation of water channels. These properties include the viscosity of the fluid source and its flow rate, both of which have not been widely investigated. 

The results on kinetics presented here are well described by a model developed by Parlange and Hill~\cite{Parlange76}, which is an extension of the model developed by Saffman and Taylor~\cite{Saffman58}. But in the application of this model, we nevertheless also take into account an additional parameter that represents the characteristic time of formation of the wetting front. The wetting front becomes homogeneous when droplet volumes at neighboring impact sites coalesce.

While perhaps this does not fully represent actual rainfall since raindrops impact randomly, it still offers an understanding of how the manner in which a fluid is injected uniformly unto the surface affects water distribution in the soil. Fluid properties and spacing between droplets influence the aggregation of droplet impact sites thus providing information that the introduction of the fluid to the porous medium is also crucial to the establishment of a wetting front.

Moreover, results show that instabilities at the wetting front and thus formation of water channels initially decreases with flow rate, particularly for low flow rates. However, depending on the viscosity, the behavior may change at larger flow rates. At low viscosities, water channels form quicker at larger flow rates. But at higher viscosities, the time of formation of channels gradually increases at higher flow rates. While our results do not yet fully explain the exact dynamics of the instability, however it does demonstrate that, within a given cross-sectional area, the number of channels that form is a function of the fluid$^\prime$s viscosity and flow rate. In terms of velocity, water channels expectedly propagate faster in larger pore sizes at constant viscosity and flow rate. In addition, water channel velocities increase with flow rate at constant particle diameter and constant viscosity. But at constant particle diameter and constant flow rate, low viscosity fluids propagate faster than larger viscosity fluids. 

We believe continuous investigations primarily focusing on the finger instability dynamics at the wetting front will further help bring to light certain aspects that remain unclear such as how the instability develops and in which particular part of the front does it develop. These experimental results presented in this paper could also be used to test existing models particularly to confirm data at conditions when the flow rate value approaches the value of the hydraulic conductivity. This typically happens when using high flow rates and high viscosity fluids. It is therefore our interest to apply these experimental results to other models of unstable multiphase flow proposed in literature with more recent ones such as the model of Cueto-Felgueroso and Juanes~\cite{Cueto09} for example. Nevertheless, apart from contributing to the advancement of our understanding of the subject, these fundamental results could also provide insights into developing techniques and applications to better control the drainage of water via preferential water channels.

\begin{acknowledgments}
We thank Jean-Christophe Castaing, Zhiyun Chen, and Larry Hough for helpful discussions. This work is financially supported by Solvay, CNRS, and by the National Science Foundation through grants MRSEC / DMR-1120901 (Y.W. and D.J.D.) and DMR-1305199 (D.J.D).
\end{acknowledgments}


\bibliography{Kinetics_References}

\begin{thebibliography}{42}
\expandafter\ifx\csname natexlab\endcsname\relax\def\natexlab#1{#1}\fi
\expandafter\ifx\csname bibnamefont\endcsname\relax
  \def\bibnamefont#1{#1}\fi
\expandafter\ifx\csname bibfnamefont\endcsname\relax
  \def\bibfnamefont#1{#1}\fi
\expandafter\ifx\csname citenamefont\endcsname\relax
  \def\citenamefont#1{#1}\fi
\expandafter\ifx\csname url\endcsname\relax
  \def\url#1{\texttt{#1}}\fi
\expandafter\ifx\csname urlprefix\endcsname\relax\def\urlprefix{URL }\fi
\providecommand{\bibinfo}[2]{#2}
\providecommand{\eprint}[2][]{\url{#2}}

\bibitem[{\citenamefont{de~Rooij}(2000)}]{DeRooij00}
\bibinfo{author}{\bibfnamefont{G.}~\bibnamefont{de~Rooij}},
  \bibinfo{journal}{Journal of Hydrology} \textbf{\bibinfo{volume}{231}},
  \bibinfo{pages}{277} (\bibinfo{year}{2000}).

\bibitem[{\citenamefont{Glass et~al.}(1991)\citenamefont{Glass, Parlange, and
  Steenhuis}}]{Glass91}
\bibinfo{author}{\bibfnamefont{R.~J.} \bibnamefont{Glass}},
  \bibinfo{author}{\bibfnamefont{J.~Y.} \bibnamefont{Parlange}},
  \bibnamefont{and} \bibinfo{author}{\bibfnamefont{T.~S.}
  \bibnamefont{Steenhuis}}, \bibinfo{journal}{Water Resources Research}
  \textbf{\bibinfo{volume}{27}}, \bibinfo{pages}{1947} (\bibinfo{year}{1991}).

\bibitem[{\citenamefont{Wang et~al.}(1998)\citenamefont{Wang, Feyen, and
  Elrick}}]{Wang98}
\bibinfo{author}{\bibfnamefont{Z.}~\bibnamefont{Wang}},
  \bibinfo{author}{\bibfnamefont{J.}~\bibnamefont{Feyen}}, \bibnamefont{and}
  \bibinfo{author}{\bibfnamefont{D.}~\bibnamefont{Elrick}},
  \bibinfo{journal}{Water Resources Research} \textbf{\bibinfo{volume}{34}},
  \bibinfo{pages}{2183} (\bibinfo{year}{1998}).

\bibitem[{\citenamefont{Hillel}(2004)}]{Hillel03}
\bibinfo{author}{\bibfnamefont{D.}~\bibnamefont{Hillel}},
  \emph{\bibinfo{title}{Introduction to Environmental Physics}}
  (\bibinfo{publisher}{Elsevier Academic Press}, \bibinfo{address}{San Diego,
  California, USA}, \bibinfo{year}{2004}).

\bibitem[{\citenamefont{Hill and Parlange}(1972)}]{Hill72}
\bibinfo{author}{\bibfnamefont{D.~E.} \bibnamefont{Hill}} \bibnamefont{and}
  \bibinfo{author}{\bibfnamefont{J.~Y.} \bibnamefont{Parlange}},
  \bibinfo{journal}{Soil Science Society of America Proceedings}
  \textbf{\bibinfo{volume}{36}}, \bibinfo{pages}{697} (\bibinfo{year}{1972}).

\bibitem[{\citenamefont{Diment and Watson}(1985)}]{Diment85}
\bibinfo{author}{\bibfnamefont{G.~A.} \bibnamefont{Diment}} \bibnamefont{and}
  \bibinfo{author}{\bibfnamefont{K.~K.} \bibnamefont{Watson}},
  \bibinfo{journal}{Water Resources Research} \textbf{\bibinfo{volume}{21}},
  \bibinfo{pages}{979} (\bibinfo{year}{1985}).

\bibitem[{\citenamefont{Ritsema et~al.}(1998)\citenamefont{Ritsema, Dekker,
  Nieber, and Steenhuis}}]{Ritsema98}
\bibinfo{author}{\bibfnamefont{C.~J.} \bibnamefont{Ritsema}},
  \bibinfo{author}{\bibfnamefont{L.~W.} \bibnamefont{Dekker}},
  \bibinfo{author}{\bibfnamefont{J.~L.} \bibnamefont{Nieber}},
  \bibnamefont{and} \bibinfo{author}{\bibfnamefont{T.~S.}
  \bibnamefont{Steenhuis}}, \bibinfo{journal}{Water Resources Research}
  \textbf{\bibinfo{volume}{34}}, \bibinfo{pages}{555} (\bibinfo{year}{1998}).

\bibitem[{\citenamefont{Hendrickx et~al.}(1993)\citenamefont{Hendrickx, Dekker,
  and Boersma}}]{Hendrickx93}
\bibinfo{author}{\bibfnamefont{J.~M.~H.} \bibnamefont{Hendrickx}},
  \bibinfo{author}{\bibfnamefont{L.~W.} \bibnamefont{Dekker}},
  \bibnamefont{and} \bibinfo{author}{\bibfnamefont{O.~H.}
  \bibnamefont{Boersma}}, \bibinfo{journal}{Journal of Environmental Quality}
  \textbf{\bibinfo{volume}{22}}, \bibinfo{pages}{109} (\bibinfo{year}{1993}).

\bibitem[{\citenamefont{Davidson}(1984)}]{Davidson84}
\bibinfo{author}{\bibfnamefont{M.}~\bibnamefont{Davidson}},
  \bibinfo{journal}{Water Resources Research} \textbf{\bibinfo{volume}{20}},
  \bibinfo{pages}{1685} (\bibinfo{year}{1984}).

\bibitem[{\citenamefont{Glass et~al.}(1989)\citenamefont{Glass, Steenhuis, and
  Parlange}}]{Glass89}
\bibinfo{author}{\bibfnamefont{R.~J.} \bibnamefont{Glass}},
  \bibinfo{author}{\bibfnamefont{T.~S.} \bibnamefont{Steenhuis}},
  \bibnamefont{and} \bibinfo{author}{\bibfnamefont{J.~Y.}
  \bibnamefont{Parlange}}, \bibinfo{journal}{Water Resources Research}
  \textbf{\bibinfo{volume}{25}}, \bibinfo{pages}{1195} (\bibinfo{year}{1989}).

\bibitem[{\citenamefont{Baker and Hillel}(1990)}]{Baker90}
\bibinfo{author}{\bibfnamefont{R.}~\bibnamefont{Baker}} \bibnamefont{and}
  \bibinfo{author}{\bibfnamefont{D.}~\bibnamefont{Hillel}},
  \bibinfo{journal}{Soil Science Society of America Journal}
  \textbf{\bibinfo{volume}{54}}, \bibinfo{pages}{20} (\bibinfo{year}{1990}).

\bibitem[{\citenamefont{Raats}(1973)}]{Raats73}
\bibinfo{author}{\bibfnamefont{P.}~\bibnamefont{Raats}}, \bibinfo{journal}{Soil
  Science Society of America Proceedings} \textbf{\bibinfo{volume}{37}},
  \bibinfo{pages}{681} (\bibinfo{year}{1973}).

\bibitem[{\citenamefont{Vanommen et~al.}(1989)\citenamefont{Vanommen, Dijksma,
  Hendrickx, Dekker, Hulshof, and Vandenheuvel}}]{Vanommen89}
\bibinfo{author}{\bibfnamefont{H.~C.} \bibnamefont{Vanommen}},
  \bibinfo{author}{\bibfnamefont{R.}~\bibnamefont{Dijksma}},
  \bibinfo{author}{\bibfnamefont{J.~M.~H.} \bibnamefont{Hendrickx}},
  \bibinfo{author}{\bibfnamefont{L.~W.} \bibnamefont{Dekker}},
  \bibinfo{author}{\bibfnamefont{J.}~\bibnamefont{Hulshof}}, \bibnamefont{and}
  \bibinfo{author}{\bibfnamefont{M.}~\bibnamefont{Vandenheuvel}},
  \bibinfo{journal}{Journal of Hydrology} \textbf{\bibinfo{volume}{105}},
  \bibinfo{pages}{253} (\bibinfo{year}{1989}).

\bibitem[{\citenamefont{Ritsema et~al.}(1993)\citenamefont{Ritsema, Dekker,
  Hendrickx, and Hamminga}}]{Ritsema93}
\bibinfo{author}{\bibfnamefont{C.~J.} \bibnamefont{Ritsema}},
  \bibinfo{author}{\bibfnamefont{L.~W.} \bibnamefont{Dekker}},
  \bibinfo{author}{\bibfnamefont{J.~M.~H.} \bibnamefont{Hendrickx}},
  \bibnamefont{and} \bibinfo{author}{\bibfnamefont{W.}~\bibnamefont{Hamminga}},
  \bibinfo{journal}{Water Resources Research} \textbf{\bibinfo{volume}{29}},
  \bibinfo{pages}{2183} (\bibinfo{year}{1993}).

\bibitem[{\citenamefont{Yao}(1993)}]{Yao93}
\bibinfo{author}{\bibfnamefont{T.~M.} \bibnamefont{Yao}}, Ph.D. thesis,
  \bibinfo{school}{New Mexico Institute of Mining and Technology}
  (\bibinfo{year}{1993}).

\bibitem[{\citenamefont{Bond}(1969)}]{Bond69}
\bibinfo{author}{\bibfnamefont{R.}~\bibnamefont{Bond}}, in
  \emph{\bibinfo{booktitle}{Proceedings on the Symposium of Water Repellent
  Soils}}, edited by \bibinfo{editor}{\bibfnamefont{L.}~\bibnamefont{DeBano}}
  \bibnamefont{and} \bibinfo{editor}{\bibfnamefont{J.}~\bibnamefont{Letey}},
  \bibinfo{organization}{Dry Lands Research Institute}
  (\bibinfo{publisher}{University of California - Riverside},
  \bibinfo{year}{1969}), pp. \bibinfo{pages}{259--264}.

\bibitem[{\citenamefont{Wang et~al.}(2000)\citenamefont{Wang, Wu, Wu, Ritsema,
  Dekker, and Feyen}}]{Wang00}
\bibinfo{author}{\bibfnamefont{Z.}~\bibnamefont{Wang}},
  \bibinfo{author}{\bibfnamefont{Q.}~\bibnamefont{Wu}},
  \bibinfo{author}{\bibfnamefont{L.}~\bibnamefont{Wu}},
  \bibinfo{author}{\bibfnamefont{C.~J.} \bibnamefont{Ritsema}},
  \bibinfo{author}{\bibfnamefont{L.~W.} \bibnamefont{Dekker}},
  \bibnamefont{and} \bibinfo{author}{\bibfnamefont{J.}~\bibnamefont{Feyen}},
  \bibinfo{journal}{Journal of Hydrology} \textbf{\bibinfo{volume}{231}},
  \bibinfo{pages}{265} (\bibinfo{year}{2000}).

\bibitem[{\citenamefont{Cejas}(2013)}]{Cejas13}
\bibinfo{author}{\bibfnamefont{C.}~\bibnamefont{Cejas}}, Ph.D. thesis,
  \bibinfo{school}{Universit{\'e} Pierre et Marie Curie - Paris 6},
  \bibinfo{address}{France} (\bibinfo{year}{2013}).

\bibitem[{\citenamefont{Green and Ampt}(1911)}]{Green11}
\bibinfo{author}{\bibfnamefont{W.}~\bibnamefont{Green}} \bibnamefont{and}
  \bibinfo{author}{\bibfnamefont{G.}~\bibnamefont{Ampt}}, \bibinfo{journal}{J.
  Agri. Sci.} \textbf{\bibinfo{volume}{4}}, \bibinfo{pages}{1}
  (\bibinfo{year}{1911}).

\bibitem[{\citenamefont{Horton}(1940)}]{Horton40}
\bibinfo{author}{\bibfnamefont{R.}~\bibnamefont{Horton}},
  \bibinfo{journal}{Soil Science Society of America Journal}
  \textbf{\bibinfo{volume}{5}}, \bibinfo{pages}{399} (\bibinfo{year}{1940}).

\bibitem[{\citenamefont{Chen and Young}(2006)}]{Chen06}
\bibinfo{author}{\bibfnamefont{L.}~\bibnamefont{Chen}} \bibnamefont{and}
  \bibinfo{author}{\bibfnamefont{M.}~\bibnamefont{Young}},
  \bibinfo{journal}{Water Resources Research} \textbf{\bibinfo{volume}{42}},
  \bibinfo{pages}{W07420} (\bibinfo{year}{2006}).

\bibitem[{\citenamefont{Brutsaert}(1977)}]{Brutsaert77}
\bibinfo{author}{\bibfnamefont{W.}~\bibnamefont{Brutsaert}},
  \bibinfo{journal}{Water Resources Research} \textbf{\bibinfo{volume}{13}},
  \bibinfo{pages}{363} (\bibinfo{year}{1977}).

\bibitem[{\citenamefont{Pachepsky et~al.}(2003)\citenamefont{Pachepsky, Timlin,
  and Rawls}}]{Pachepsky03}
\bibinfo{author}{\bibfnamefont{Y.}~\bibnamefont{Pachepsky}},
  \bibinfo{author}{\bibfnamefont{D.}~\bibnamefont{Timlin}}, \bibnamefont{and}
  \bibinfo{author}{\bibfnamefont{W.}~\bibnamefont{Rawls}},
  \bibinfo{journal}{Journal of Hydrology} \textbf{\bibinfo{volume}{272}},
  \bibinfo{pages}{3} (\bibinfo{year}{2003}).

\bibitem[{\citenamefont{Parlange and Hill}(1976)}]{Parlange76}
\bibinfo{author}{\bibfnamefont{J.~Y.} \bibnamefont{Parlange}} \bibnamefont{and}
  \bibinfo{author}{\bibfnamefont{D.~E.} \bibnamefont{Hill}},
  \bibinfo{journal}{Soil Science} \textbf{\bibinfo{volume}{122}},
  \bibinfo{pages}{236} (\bibinfo{year}{1976}).

\bibitem[{\citenamefont{Cueto-Felgueroso and
  Juanes}(2008)}]{Cueto-Felgueroso08}
\bibinfo{author}{\bibfnamefont{L.}~\bibnamefont{Cueto-Felgueroso}}
  \bibnamefont{and} \bibinfo{author}{\bibfnamefont{R.}~\bibnamefont{Juanes}},
  \bibinfo{journal}{Physical Review Letters} \textbf{\bibinfo{volume}{101}},
  \bibinfo{pages}{244504} (\bibinfo{year}{2008}).

\bibitem[{\citenamefont{Cueto-Felgueroso and
  Juanes}(2009{\natexlab{a}})}]{Cueto-Felgueroso09}
\bibinfo{author}{\bibfnamefont{L.}~\bibnamefont{Cueto-Felgueroso}}
  \bibnamefont{and} \bibinfo{author}{\bibfnamefont{R.}~\bibnamefont{Juanes}},
  \bibinfo{journal}{Physical Review E} \textbf{\bibinfo{volume}{79}},
  \bibinfo{pages}{036301} (\bibinfo{year}{2009}{\natexlab{a}}).

\bibitem[{\citenamefont{Nieber et~al.}(2005)\citenamefont{Nieber, Duatov,
  Egorov, and Sheshukov}}]{Nieber05}
\bibinfo{author}{\bibfnamefont{J.~L.} \bibnamefont{Nieber}},
  \bibinfo{author}{\bibfnamefont{R.}~\bibnamefont{Duatov}},
  \bibinfo{author}{\bibfnamefont{A.}~\bibnamefont{Egorov}}, \bibnamefont{and}
  \bibinfo{author}{\bibfnamefont{A.}~\bibnamefont{Sheshukov}},
  \bibinfo{journal}{Transp. Porous Media} \textbf{\bibinfo{volume}{58}},
  \bibinfo{pages}{147} (\bibinfo{year}{2005}).

\bibitem[{\citenamefont{Eliassi and Glass}(2001)}]{Eliassi01}
\bibinfo{author}{\bibfnamefont{M.}~\bibnamefont{Eliassi}} \bibnamefont{and}
  \bibinfo{author}{\bibfnamefont{R.~J.} \bibnamefont{Glass}},
  \bibinfo{journal}{Water Resources Research} \textbf{\bibinfo{volume}{37}},
  \bibinfo{pages}{2019} (\bibinfo{year}{2001}).

\bibitem[{\citenamefont{Saffman and Taylor}(1958)}]{Saffman58}
\bibinfo{author}{\bibfnamefont{P.}~\bibnamefont{Saffman}} \bibnamefont{and}
  \bibinfo{author}{\bibfnamefont{G.~S.} \bibnamefont{Taylor}},
  \bibinfo{journal}{Proc. Roy. Soc. London A} \textbf{\bibinfo{volume}{245}},
  \bibinfo{pages}{312} (\bibinfo{year}{1958}).

\bibitem[{\citenamefont{Chuoke et~al.}(1959)\citenamefont{Chuoke, Vanmeurs, and
  Vanderpoel}}]{Chuoke59}
\bibinfo{author}{\bibfnamefont{R.~L.} \bibnamefont{Chuoke}},
  \bibinfo{author}{\bibfnamefont{P.}~\bibnamefont{Vanmeurs}}, \bibnamefont{and}
  \bibinfo{author}{\bibfnamefont{C.}~\bibnamefont{Vanderpoel}},
  \bibinfo{journal}{Transactions of the American Institute of Mining and
  Metallurgical Engineers} \textbf{\bibinfo{volume}{216}}, \bibinfo{pages}{188}
  (\bibinfo{year}{1959}).

\bibitem[{\citenamefont{Philip}(1969)}]{Philip69}
\bibinfo{author}{\bibfnamefont{J.}~\bibnamefont{Philip}},
  \bibinfo{journal}{Adv. Hydrosci.} \textbf{\bibinfo{volume}{5}},
  \bibinfo{pages}{215} (\bibinfo{year}{1969}).

\bibitem[{\citenamefont{DiCarlo and Blunt}(2000)}]{DiCarlo00}
\bibinfo{author}{\bibfnamefont{D.}~\bibnamefont{DiCarlo}} \bibnamefont{and}
  \bibinfo{author}{\bibfnamefont{M.}~\bibnamefont{Blunt}},
  \bibinfo{journal}{Water Resources Research} \textbf{\bibinfo{volume}{36}},
  \bibinfo{pages}{2781} (\bibinfo{year}{2000}).

\bibitem[{\citenamefont{Cheng}(2000)}]{Cheng00}
\bibinfo{author}{\bibfnamefont{N.}~\bibnamefont{Cheng}}, \bibinfo{journal}{Ind.
  Eng. Chem. Res.} \textbf{\bibinfo{volume}{47}}, \bibinfo{pages}{3285}
  (\bibinfo{year}{2000}).

\bibitem[{\citenamefont{Shchekotov}(2010)}]{Shchekotov10}
\bibinfo{author}{\bibfnamefont{D.}~\bibnamefont{Shchekotov}}, Master's thesis,
  \bibinfo{school}{University of Stavanger}, \bibinfo{address}{Norway}
  (\bibinfo{year}{2010}).

\bibitem[{\citenamefont{Drelich et~al.}(1996)\citenamefont{Drelich, Wilbur, and
  Whitesides}}]{Drelich96}
\bibinfo{author}{\bibfnamefont{J.}~\bibnamefont{Drelich}},
  \bibinfo{author}{\bibfnamefont{J.}~\bibnamefont{Wilbur}}, \bibnamefont{and}
  \bibinfo{author}{\bibfnamefont{G.}~\bibnamefont{Whitesides}},
  \bibinfo{journal}{Langmuir} \textbf{\bibinfo{volume}{12}},
  \bibinfo{pages}{1913} (\bibinfo{year}{1996}).

\bibitem[{\citenamefont{Wei et~al.}(2014)\citenamefont{Wei, Cejas, Barrois,
  Dreyfus, and Durian}}]{Wei13}
\bibinfo{author}{\bibfnamefont{Y.}~\bibnamefont{Wei}},
  \bibinfo{author}{\bibfnamefont{C.}~\bibnamefont{Cejas}},
  \bibinfo{author}{\bibfnamefont{R.}~\bibnamefont{Barrois}},
  \bibinfo{author}{\bibfnamefont{R.}~\bibnamefont{Dreyfus}}, \bibnamefont{and}
  \bibinfo{author}{\bibfnamefont{D.~J.} \bibnamefont{Durian}},
  \emph{\bibinfo{title}{Morphology of rain water channelization in
  systematically varied model sandy soils}} (\bibinfo{year}{2014}),
  \bibinfo{note}{"Morphology of rain water channelization in systematically
  varied model sandy soils" (submitted with this article)}.

\bibitem[{\citenamefont{Culligan et~al.}(2005)\citenamefont{Culligan, Ivanov,
  and Germaine}}]{Culligan05}
\bibinfo{author}{\bibfnamefont{P.~J.} \bibnamefont{Culligan}},
  \bibinfo{author}{\bibfnamefont{V.}~\bibnamefont{Ivanov}}, \bibnamefont{and}
  \bibinfo{author}{\bibfnamefont{J.~T.} \bibnamefont{Germaine}},
  \bibinfo{journal}{Advances in Water Resources} \textbf{\bibinfo{volume}{28}},
  \bibinfo{pages}{1010} (\bibinfo{year}{2005}).

\bibitem[{\citenamefont{Rezanezhad}(2007)}]{Rezanezhad07}
\bibinfo{author}{\bibfnamefont{F.}~\bibnamefont{Rezanezhad}}, Ph.D. thesis,
  \bibinfo{school}{University of Heidelberg} (\bibinfo{year}{2007}).

\bibitem[{\citenamefont{Cueto-Felgueroso and
  Juanes}(2009{\natexlab{b}})}]{Cueto09}
\bibinfo{author}{\bibfnamefont{L.}~\bibnamefont{Cueto-Felgueroso}}
  \bibnamefont{and} \bibinfo{author}{\bibfnamefont{R.}~\bibnamefont{Juanes}},
  \bibinfo{journal}{Water Resour. Res.} \textbf{\bibinfo{volume}{45}},
  \bibinfo{pages}{W10409} (\bibinfo{year}{2009}{\natexlab{b}}).

\bibitem[{\citenamefont{Cho and de~Rooij}(2002)}]{Cho02}
\bibinfo{author}{\bibfnamefont{H.}~\bibnamefont{Cho}} \bibnamefont{and}
  \bibinfo{author}{\bibfnamefont{G.}~\bibnamefont{de~Rooij}},
  \bibinfo{journal}{Hydrology and Earth System Sciences}
  \textbf{\bibinfo{volume}{6}}, \bibinfo{pages}{763} (\bibinfo{year}{2002}).

\bibitem[{\citenamefont{Verneuil and Durian}(2011)}]{Verneuil11}
\bibinfo{author}{\bibfnamefont{E.}~\bibnamefont{Verneuil}} \bibnamefont{and}
  \bibinfo{author}{\bibfnamefont{D.~J.} \bibnamefont{Durian}},
  \bibinfo{journal}{European Physical Journal E} \textbf{\bibinfo{volume}{34}},
  \bibinfo{pages}{65} (\bibinfo{year}{2011}).

\bibitem[{\citenamefont{Selker et~al.}(1992)\citenamefont{Selker, Steenhuis,
  and Parlange}}]{Selker92}
\bibinfo{author}{\bibfnamefont{J.~S.} \bibnamefont{Selker}},
  \bibinfo{author}{\bibfnamefont{T.~S.} \bibnamefont{Steenhuis}},
  \bibnamefont{and} \bibinfo{author}{\bibfnamefont{J.~Y.}
  \bibnamefont{Parlange}}, \bibinfo{journal}{Soil Science Society of America
  Journal} \textbf{\bibinfo{volume}{56}}, \bibinfo{pages}{1346}
  (\bibinfo{year}{1992}).

\end{thebibliography}


%
\end{document}